\documentclass[aps,pra,letterpaper,superscriptaddress,twocolumn,floatfix]{revtex4-2}
\usepackage{graphicx}
\usepackage{dcolumn}

\usepackage{bm}

\usepackage{subfigure}
\usepackage{xcolor}
\newcommand{\be}{\begin{eqnarray}}
\newcommand{\ee}{\end{eqnarray}}
\usepackage{amsfonts}
\usepackage{bbm}
\usepackage{amsmath,leftidx}
\usepackage{graphicx}
\usepackage{times}
\usepackage{CJK}
\usepackage{color,colortbl}
\usepackage[normalem]{ulem}
\usepackage{hyperref}

\begin{document}

\title{
Detecting  transition between  Abelian and non-Abelian topological orders \\ through symmetric tensor networks
}

\author{ Yu-Hsueh Chen}
\affiliation{ Department of Physics and Center for Theoretical Physics, National Taiwan University, Taipei 10607, Taiwan }

\author{ Ching-Yu Huang }
\email{cyhuangphy@thu.edu.tw}
\affiliation{ Department of Applied Physics, Tunghai University, Taichung 40704, Taiwan}

\author{ Ying-Jer Kao }
\email{yjkao@phys.ntu.edu.tw}
\affiliation{ Department of Physics and Center for Theoretical Physics, National Taiwan University, Taipei 10607, Taiwan }

\begin{abstract}

We propose a unified scheme to identify  phase transitions out of the  $\mathbb{Z}_2$ Abelian topological order, including the transition to a non-Abelian chiral spin liquid. 
Using  loop gas  and and string gas states [H.-Y. Lee, R. Kaneko, T. Okubo, N. Kawashima, Phys. Rev. Lett. 123, 087203 (2019)] on the star lattice Kitaev model as an example, we compute the overlap of minimally entangled states through transfer matrices.
We demonstrate that, similar to the anyon condensation, continuous deformation of a $\mathbb{Z}_2$-injective projected entangled-pair  state (PEPS) also allows us to study the transition between  Abelian and non-Abelian topological orders. 
We  show that the  charge and flux anyons  defined in the Abelian phase transmute into the $\sigma$ anyon in the non-Abelian topological order.
Furthermore, we  show that contrary to the claim in  Phys. Rev. B 101, 035140 (2020),  both the LG and SG states have infinite correlation length in the non-Abelian regime, consistent with the no-go theorem that a chiral PEPS  has a gapless parent Hamiltonian.

\end{abstract}

\maketitle


\section{ Introduction}

In the past decades, significant efforts have been devoted to understanding topologically ordered phases and topological phase transitions. 
Topological phases~\cite{Wen1989,Wen1990} cannot be characterized by a local order parameter and can be characterized by properties such as the ground-state degeneracy, and anyonic quasiparticle statistics~\cite{wen_1990,Wen1993,KITAEV20062, Bais_2012, Zhang_2012}.
Recently, it was realized that these states can be understood using notions such as topological entanglement entropy~\cite{ Kitaev2006, Levin2006,Xie_2010}, and  entanglement spectrum~\cite{Haldane_2008,Pollmann_2010,Turner_2011}.
The latter reveals the edge physics of the topological state, and can be easily computed for the projected entangled-pair states (PEPS)~\cite{PEPS}, a type of tensor networks that has been successfully represented the ground state wavefunction for systems with both conventional and topological orders. 
For a  symmetry group $G$, the $G$-injective PEPSs ~\cite{2011_Norbert_Ginjective} form a natural framework to describe the anyon theory given by the quantum double $D(G)$. 
It  encodes topological properties in the local symmetries on the virtual indices. 
A $G$-injective PEPS encodes the ground state subspace of its parent Hamiltonian, allowing us to study the topological properties through its entanglement degrees of freedom. 
However, it was shown that a $G$-injective PEPS does not guarantee a topologically ordered phase since  the system can be driven into a topologically trivial phase by a physical deformation of the local tensor~\cite{Norbert_Schuch_2013,Haegeman_2015, 2017-PRB-Norbert-Entanglement, 2017_Z4_anyon, 2017_sym_induced,2018_Chen_Boson_condensation, Zhang_2019, PhysRevB.101.041108}. 
For example, the phases and phase transitions of the two dimensional (2D) toric code (TC) model with finite string tension, whose ground state is represented by  a $\mathbb{Z}_2$-injective  PEPS~\cite{2017-PRB-Norbert-Entanglement},  can be fully understood within this framework. 
Similar idea of detecting topological phase transitions has also been generalized to non-Abelian cases recently~\citep{PhysRevB.96.155127, PhysRevB.100.245125, PhysRevLett.124.130603, PhysRevB.102.235112}. 

A {new} class of $\mathbb{Z}_2 $-injective ansatz called loop gas (LG) and string gas (SG)  is constructed to represent the ground state of the  Kitaev models~\cite{KITAEV20062} on the honeycomb lattice~\cite{spin_one_half,2020-spin-one-kitaev,lee2020anisotropy}. 
Surprisingly,  when the same  ansatz is applied to the Kitaev model on the star lattice,  the entanglement entropy and  spectrum suggest that flux anyon in the $\mathbb{Z}_2 $-topological order become the $\sigma$ anyon in the non-Abelian chiral spin liquid~(CSL)~\cite{non-AbelianTO_2020}. 
However, exact results show that the ground state subspace should be three dimensional  in the non-Abelian regime~\cite{Hong_Ya_2007},  inconsistent with the four-fold degenerate ground state structure for the $\mathbb{Z}_2 $-injective PEPS. 
In this paper,  we propose to use the overlap of minimally entangled states (MESs) ~\cite{2012-PRB-Oshikawa-MES} as a unified framework to understand  the phase transitions out of a $\mathbb{Z}_2$ topological order. 
 By computing the transfer matrices (TMs) associating with MESs on the long cylinder, the overlap can be obtained from the dominant eigenvalue of the corresponding TMs.
Our results show that similar to anyon condensation, the transition from an Abelian to a non-Abelian topologically ordered phase can be understood as both the charge and flux transmute into the $\sigma$ anyon, resolving the mismatch between the dimension of the ground state subspace for a $\mathbb{Z}_2 $-injective PEPS and degeneracy of the CSL ground state of the star lattice Kitaev model. 
We also show that the correlation lengths of LG and SG states diverge in the CSL regime, consistent with the recent claim that the parent Hamiltonian of a chiral PEPS is gapless~\cite{2015_no_go_theorem,2013_PEPS_chiral_TO,2015_chiral_PEPS_TO}.

This paper is organized as follows,
In Sec.~\ref{sec:Z2-injective}, we briefly review the properties of $\mathbb{Z}_2 $-injective PEPS.
In Sec.~\ref{sec:MES}, we show  the relation between  the overlap of MESs and the transfer matrix. 
In Sec.~\ref{sec:Toric Code}, we revisit the toric code with string tension and demonstrate how to detect anyon condensation transitions using the MES overlap picture. 
In Sec.~\ref{sec:LG/ SG state },  we apply the method to the Kitaev model on the star lattice and show how to detect the Abelian to non-Abelian topological order transition using MES overlap. 
We show that the flux and charge anyons transmute into the $\sigma$ anyon from the full transfer matrix spectrum.
We conclude in Sec.~\ref{sec:Conclusion}.

\section{\label{sec:Z2-injective}Symmetric PEPS and anyons}

\begin{figure}[t]
\centering
\includegraphics[width=\linewidth]{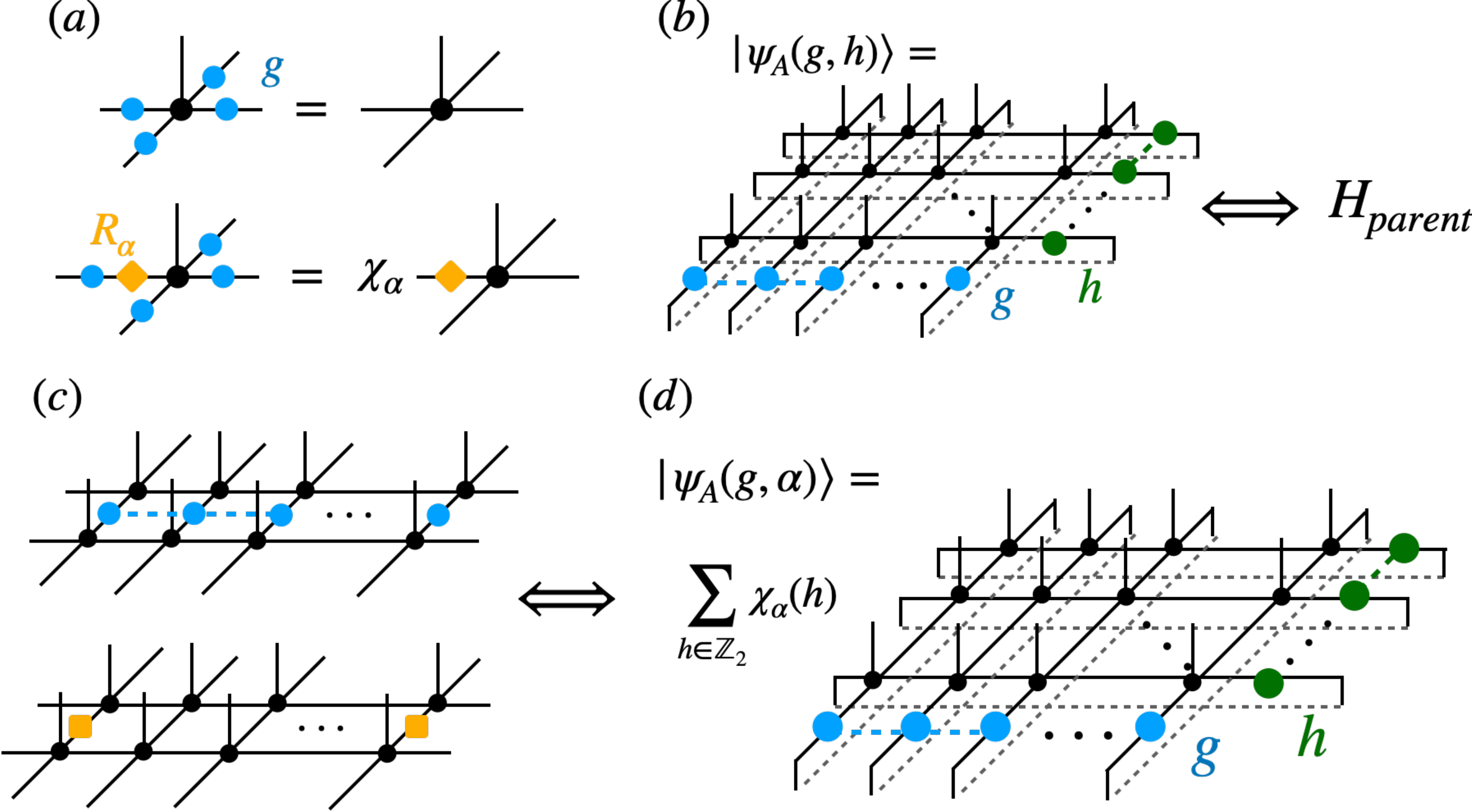}
\caption{(a) A $\mathbb{Z}_2$-injective PEPS is invariant under $A(u_g \otimes u_g \otimes u_g^\dagger \otimes u_g^\dagger) = A $, where $g \in \{{I}, {Z}\}$. 
On the other hand, we can find an operator $R_\alpha$ such that it transforms non-trivially under the group action. (b) For any $\mathbb{Z}_2$-injective PEPS $A$, we can construct a parent Hamiltonian such that its ground state subspace on the torus is spanned by $|\psi_A(g,h)\rangle, \,  \forall g,h \in \{{I}, Z\}$  (c) The anyon excitation can be constructed by either attaching a string of $u_g$ (flux) or applying $R_\alpha$ (charge) on the virtual indices. (d) The MES basis relating to anyon excitations can be constructed through $|\psi_A(g, \alpha)\rangle = \sum_{h \in \mathbb{Z}_2 }\chi_\alpha(h) |\psi_A(g,h) \rangle$.}
\label{Z2_PEPS}
\end{figure}

A translational invariant PEPS wave function can be  written in terms of a local tensor $A_{\alpha\beta\gamma\delta}^i$ with the physical index $i$ and virtual indices $\alpha,\beta,\gamma,\delta$ as
\begin{equation}
\left|\psi_A\right\rangle=\sum_{i_1,\ldots,i_N} \operatorname{tTr}\left(A^{i_1}A^{i_2}\ldots A^{i_N}\right)|i_1, i_2,\ldots,i_N\rangle,
\end{equation}
where the tensorial trace is over the virtual indices. 
A  $\mathbb{Z}_2$-invariant  PEPS [Fig.~\ref{Z2_PEPS}(a)] is represented by a local tensor $A$ that is  invariant  under the global $\mathbb{Z}_2$ symmetry, i.e., $A(u_g \otimes u_g \otimes u_g^\dagger \otimes u_g^\dagger) = A $, where $u_g$ is a representation of the group $\mathbb{Z}_2 $ with $g \in \{I, Z\}$. 
If $\mathbb{Z}_2 $ is the only symmetry of  $A$, we  say that $A$ is $\mathbb{Z}_2 $-injective ~\cite{2011_Norbert_Ginjective}.
For a $\mathbb{ Z}_2$-injective PEPS,
the ground state subspace of the parent Hamiltonian is spanned by $|\psi_A(g,h)\rangle$, corresponding to acting the two non-contractible loop operators, $(u_g^{\otimes L_x}, u_h^{\otimes L_y} ),\  \forall g,h \in \{{I}, Z\}$, on the original PEPS states [Fig.~\ref{Z2_PEPS}(b)]. {Here $L_x(L_y)$ is the system size along the $x(y)$-direction.  }
This arises from the fact that the parent Hamiltonian cannot detect these loop operators locally, as we can always deform the non-contractible loop operators  using the $\mathbb{Z}_2$-invariant property.
The $\mathbb{ Z}_2$-injective tensor naturally supports anyonic excitations that cannot be created locally on the systems. 
For example, a flux excitation can be created by attaching a string of $u_g,\ g\in \{{I}, Z\}$ on the virtual bond.
A charge excitation can be created by acting an operator $R_\alpha$ on the virtual bond which transform non-trivially under the group action $R_\alpha u_g = \chi_\alpha(g) u_g R_\alpha$, where $\chi$ is the character and $\alpha$ designates the irreducible representation of  $\mathbb{Z}_2$ [Fig.~\ref{Z2_PEPS}(c)]~\citep{2011_Norbert_Ginjective,2017-PRB-Norbert-Entanglement}.

Note that far away from the renormalization group fixed point, the excitation is dispersive and   local action of $u_g$ and $R_\alpha$ may not correspond to the eigenstates of  the parent Hamiltonian.
Instead, the excited states should be created by the superposition of local excitations. 
However, as shown in Ref.~\cite{Haegeman_2015}, these local actions remain crucial to extract anyonic information.

\section{Minimal entangled states and transfer matrices }
\label{sec:MES}
Ground states subspace and the {anyonic} excitation are closely related, and we can construct a special ground state basis, the minimally entangled states (MESs), to reflect the anyonic excitation of  the topological phases. 
Basically, the MES basis can be obtained by creating a pair of anyons on a torus, wrapping them around a closed non-contractible loop, and finally annihilating them. 
To be  specific, a MES is the eigenstate of the Wilson loop operator with a definite type of anyon excitation; therefore we can  construct the MESs in the ground state subspace by  $|\psi_A(g, \alpha)\rangle = \sum_{h \in \mathbb{Z}_2 }\chi_\alpha(h) |\psi_A(g,h) \rangle$, with $g\in \{{I}, Z\}$ [Fig.~\ref{Z2_PEPS}(d)]. 

\begin{figure}[t]
 \centering
\includegraphics[width=\linewidth]{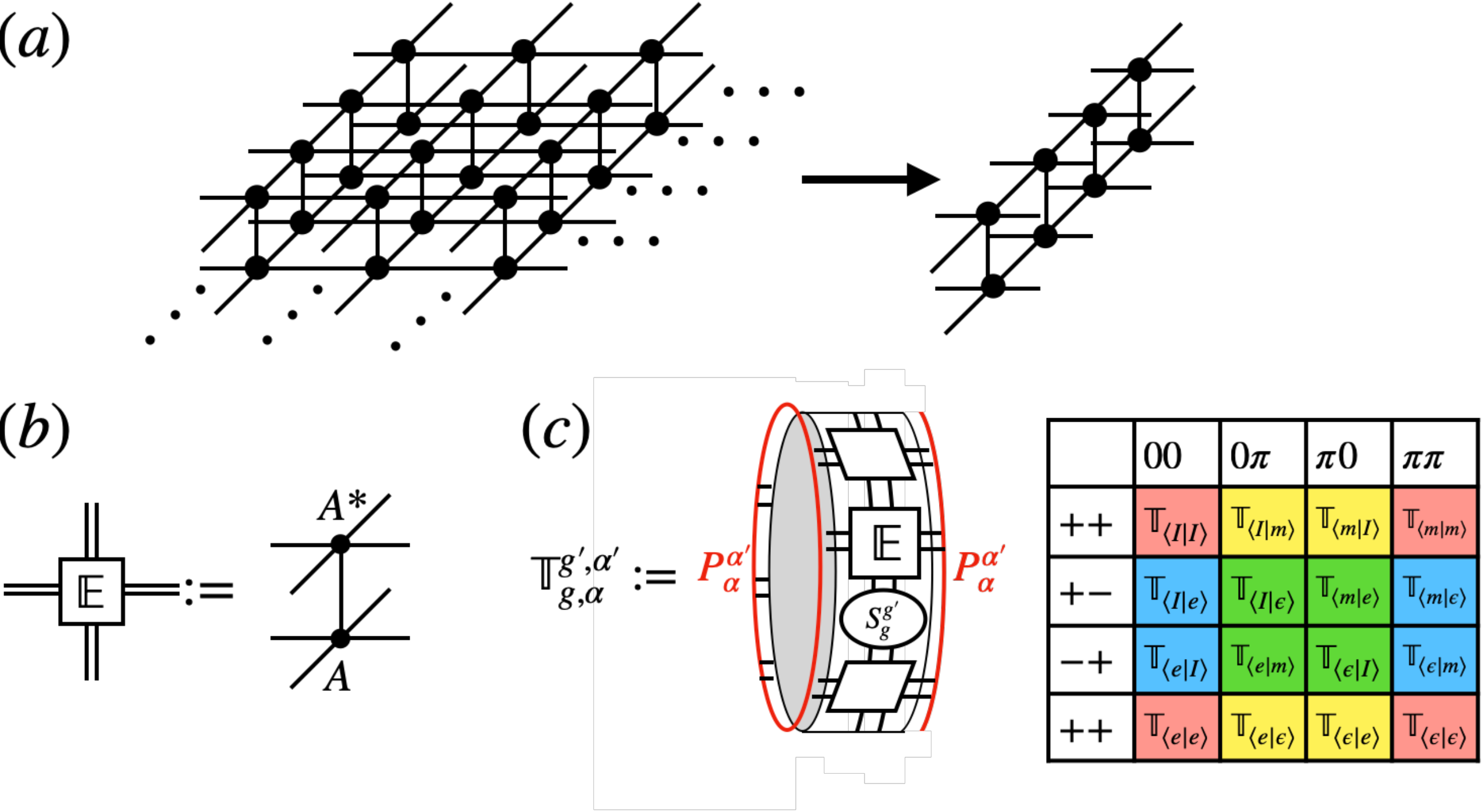}
\caption{(a) Overlap of two-dimensional infinite PEPS states can be regarded as  a one-dimensional transfer matrix. (b) A double tensor is formed by contracting  tensor $A$ on a lattice site $a$ with its adjoint tensor $A^*$ over the physical index. (c) Sixteen blocks of  the transfer matrices. 
}  
\label{fig:double_tensor}
\end{figure}

In the following, we denote $g = {I} (Z)$ as 0($\pi$)-flux and the parity $\alpha$ as $+(-)$.
The four MESs $|I\rangle ,|e \rangle, |m \rangle, |\epsilon\rangle $ then correspond to  $|\psi_A(0,+)\rangle,|\psi_A(0,-)\rangle,|\psi_A(\pi,+)\rangle,|\psi_A(\pi,-)\rangle $, respectively.
The overlap of MESs provides the information about the identities of the anyonic excitations and can be obtained from a  transfer matrix (TM) [Fig.~\ref{fig:double_tensor} (a) ].  
Starting from a local tensor $A$ representing a $\mathbb{ Z}_2 $ topological order, 
we  form  a double tensor  $\mathbb{ E} $ [Fig.~\ref{fig:double_tensor} (b)] by contracting the physical indices of   $A$ and its adjoint $A^*$,
 $\mathbb{ E}  \equiv \sum_s ( A^s_{i,j,k,l} ) \times ( A^s_{i',j',k',l'} )^* $. 
 The corresponding  transfer  matrix  is given by 
 \begin{equation}
\mathbb{ T}(L_y)  \equiv  \operatorname{tTr} ( \mathbb{ E}^1  \mathbb{ E}^2 \cdots \mathbb{ E}^{L_y}   ). 
\end{equation}
Here the tensorial trace is along the $y$-direction.
The  minimally entangled topological sectors corresponding to the quasiparticles can be obtained by inserting
 the string operator   $S^{g'}_g = u_{g'} \otimes u_g$ ($g = I,Z$) along the cylinder direction and choosing the boundary conditions $ P_{\alpha}^ {\alpha'} = P_{\alpha'} \otimes P_{\alpha}$ ($\alpha, \alpha'= +,-$).
Here $P_{\pm}$ is a projector onto the irreducible representations $ \pm 1$ of ${u_Z}^{ \otimes L_y}$.
Overall, it gives 16 blocks of  transfer matrices,
which are defined as  [Fig.~\ref{fig:double_tensor} (c)]
 \begin{equation}
\mathbb{ T}_{\langle a|b \rangle}(L_y) = \mathbb{ T}_{g,\alpha}^ {g',\alpha'} (L_y) \equiv  P_{\alpha}^ {\alpha'}  \big[ \operatorname{tTr} ( \mathbb{ E}^1  \mathbb{ E}^2 \cdots \mathbb{ E}^{L_y}  {S}_g^{g'} )  \big]  P_{\alpha}^ {\alpha'}. 
\label{eq:TM}
\end{equation}
The overlap of two MESs in the thermodynamic limit is 
 \begin{equation}
\langle a|b\rangle =  \lim_{L_x \rightarrow \infty } \operatorname{tTr}{\bigg[  \big( \lim_{L_y \rightarrow \infty} \mathbb{ T}_{\langle a|b \rangle} (L_y) \big)^{L_x} \bigg] }
\label{eq:overlap}
\end{equation}
Here, the tensorial trace is along the $x$-direction. It is then obvious that only the largest eigenvalue $ \lambda_{\langle a|b \rangle}$  of   $ \mathbb{ T}_{\langle a|b \rangle}(L_y)$ survives after the power of $L_x$ in Eq.~\eqref{eq:overlap}. Therefore, 
the leading eigenvalue of a TM gives the overlap of the two MESs in the thermodynamic limit. 
In other words, if $\lambda_{\langle a|b \rangle} = 1$, then $\langle a|b\rangle = \lim_{L_x \rightarrow \infty }\lambda_{\langle a|b \rangle}^{L_x} = 1$, while $\langle a|b\rangle = \lim_{L_x \rightarrow \infty }\lambda_{\langle a|b \rangle}^{L_x} = 0$ if $\lambda_{\langle a|b \rangle} < 1$.

 In the $\mathbb{Z}_2$ topological order phase, the transfer matrices can be divided into four types:
 (a) $\alpha = \alpha'$ and $g=g'$ corresponds to the regular TM computing the norm of the MES, ${\langle a | a \rangle}$, with $ a = I,e,m, \epsilon $ (red color in Fig.~\ref{fig:double_tensor}(c)), 
 (b) $\alpha \neq  \alpha'$ and $g=g'$ corresponds to the { mixed} TM measuring the charge difference between the bra and ket (blue), 
 (c) $\alpha =  \alpha'$ and $g \neq g'$ corresponds to the { mixed} TM measuring the flux difference between the bra and ket  (yellow), 
 (d) $\alpha \neq  \alpha'$ and $g \neq g'$ corresponds to the {mixed} TMs measuring the both charge and flux (fermion) difference between the bra and ket (green). 

Far away from the renormalization group fixed point, the subleading eigenvalues of the regular transfer matrix are related to the excitation energy, and the dominant eigenvalues of the transfer matrices measuring the charge (flux) difference  are   related to the charge (flux) excitation energy ~\citep{Zauner_2015,Haegeman_2015}. 
Similarly, the green blocks in Fig.~\ref{fig:double_tensor}(c) ($\alpha \neq  \alpha'$ and $g \neq g'$ )  correspond to the fermionic excitation. 
While the overall energy scale of eigenvalues and exact excitation energies are unknown due to the lack of Lieb-Robinson velocity \citep{1972_Lieb},  %
this  correspondence makes the relation between MES and anyon more apparent, allowing us to use these 16 transfer  matrices to study anyon condensation.
As we will discuss in the following section, by tuning the wave function without spoiling the $\mathbb{Z}_2$-injectivity,  it is possible to drive the system from a topological ordered phase to a topological trivial phase through the condensation of charge(flux) anyons.
In the MES language, this simply means that $|e\rangle (|m\rangle)$  is identical to $|I\rangle$, i.e., wrapping a charge pair around a non-contractable loop and annihilating them cannot produce a linearly independent state from the original ground state \citep{Norbert_Schuch_2013}.

It is also possible that an anyon can transmute into another type of anyon. 
For example, when the $D(\mathbb{Z}_4)$ quantum double model is continuously deformed to the toric code or the double semion model, some of the anyons distinct in the $D(\mathbb{Z}_4)$ phase can be identified as the same~\cite{2017-PRB-Norbert-Entanglement, 2017_Z4_anyon}. 
Similarly, in the case of the $\mathbb{Z}_2$-injective PEPS, we can ask which anyons we can identify as the same.  

Since the TM is periodic around  the cylinder, we can label the states with the momentum quantum number.
Interestingly, it has been shown in Ref.~\citep{Haegeman_2015} that the momentum quantum number of $|\epsilon\rangle$ will be shifted by half a spacing, i.e., $k = 2\pi(n+\frac{1}{2})/L_y$, where $n = 0,\ldots ,L_y-1$ and $L_y$ is the circumference of the cylinder. 
This momentum polarization~\citep{2013_momentum_polarization} makes $|\epsilon\rangle$ impossible to become other MESs.
This makes sense in the anyon condensation picture that fermion can never condense (corresponds to $|\epsilon \rangle \neq |I\rangle$) within the framework of  $G$-injective PEPS~\citep{2017-PRB-Norbert-Entanglement}.
Discarding the possibility that  $|\epsilon\rangle$ becomes $|e\rangle$ or $|m\rangle$  and the well studied case that $|e\rangle$ or $|m\rangle$  becoming $|I\rangle$, we are left with the only choice to identify $|e\rangle$ and $|m\rangle$ as the same state. 
Later we will see that the LG and SG states satisfy this condition. 

Remarkably, the fact that no other MESs can become $|\epsilon \rangle$ enables us to determine whether the system reach the phase transition point. 
As argued before, if $\lambda_{\langle a|b\rangle} \rightarrow 1$ as $L_y \rightarrow \infty$, it generally means that $|a\rangle = |b\rangle$. 
However, it is possible that in some situations, $\lambda_{\langle i|\epsilon \rangle} \rightarrow 1$ as $L_y \rightarrow \infty$ for $i = I,e,m$. This means that the construction of MES basis is not well-defined and the system is gapless. We will demonstrate this usage in the following sections.

In order to efficiently contract the TMs on the long cylinder, we coarse grain the tensors  along the $y$-direction using HOTRG-like method \citep{PhysRevB.86.045139}. During the coarse-graining process, the gauge symmetry is preserved to separate the 16 blocks of TM in Fig.~\ref{fig:double_tensor}(c) \cite{GSPRG_2014} (see App.~\ref{sec: tnTM} for more details).

\section{Toric Code with finite String Tension} 
\label{sec:Toric Code}
Here we revisit the toric code (TC) model with finite string tension, which is the simplest example with the phase transition from a topological order to a topologically trivial phase~\cite{Norbert_Schuch_2013,Haegeman_2015}.
We  add  the string tension   by applying the operator $Q_e(\beta_x, \beta_z) = \exp \left(  \frac{  \beta_x \sigma^x_e + \beta_z \sigma^z_e }{4}  \right)$ to the TC, 
\begin{align}
|\Psi(\beta_x,\beta_z)\rangle &= \prod_{e} Q_e(\beta_x,\beta_z)\times \nonumber\\
 &\prod_{v}\left(1+\prod_{e \ni v} \sigma_{e}^{x}\right) \prod_{p}\left(1+\prod_{e \in \partial p} \sigma_{e}^{z}\right) |\Omega\rangle,
\end{align}
where $e/p$ labels the the vertex/plaquette, and $ |\Omega\rangle$ indicates  the fully polarized spin state $|\Omega\rangle=\otimes_{e}|\uparrow\rangle_{e}$. 
For $ \beta_x =0$ and  $\beta_z \to \infty$, the system is driven to the charge condensed (CC) phase. 
On the other hand,  as $ \beta_x \to \infty $ and  $\beta_z $ = 0, the system is driven to the flux condensed (FC) phase.
Therefore, it is expected that by tuning the parameters $ \beta_x, \beta_z$, phase transitions will occur.

At the $\mathbb{ Z}_2 $ topological order (TO) fixed point, the eigenvalues $ \lambda_{\langle a | a \rangle} =1 $, $( a =I,e,m, \epsilon )$, and zero otherwise, indicating these four MESs are orthonormal. 
At the fixed point of the CC phase, $\lambda_{\langle I|I\rangle} = \lambda_{\langle I| e \rangle} = \lambda_{\langle  e| I \rangle} = \lambda_{\langle e | e \rangle}  = 1 $, and zero otherwise, suggesting that sector $|e\rangle$ is identical to $|I\rangle$ while $|m\rangle$ and $|\epsilon\rangle$ are confined.  
Similarly, at the fixed point of the FC phase, we have $\lambda_{\langle I |I\rangle} = \lambda_{\langle I| m \rangle} = \lambda_{\langle   m | I \rangle} = \lambda_{\langle m | m \rangle}  = 1$.

\begin{figure}[t]
\centering
\includegraphics[width=0.8\linewidth]{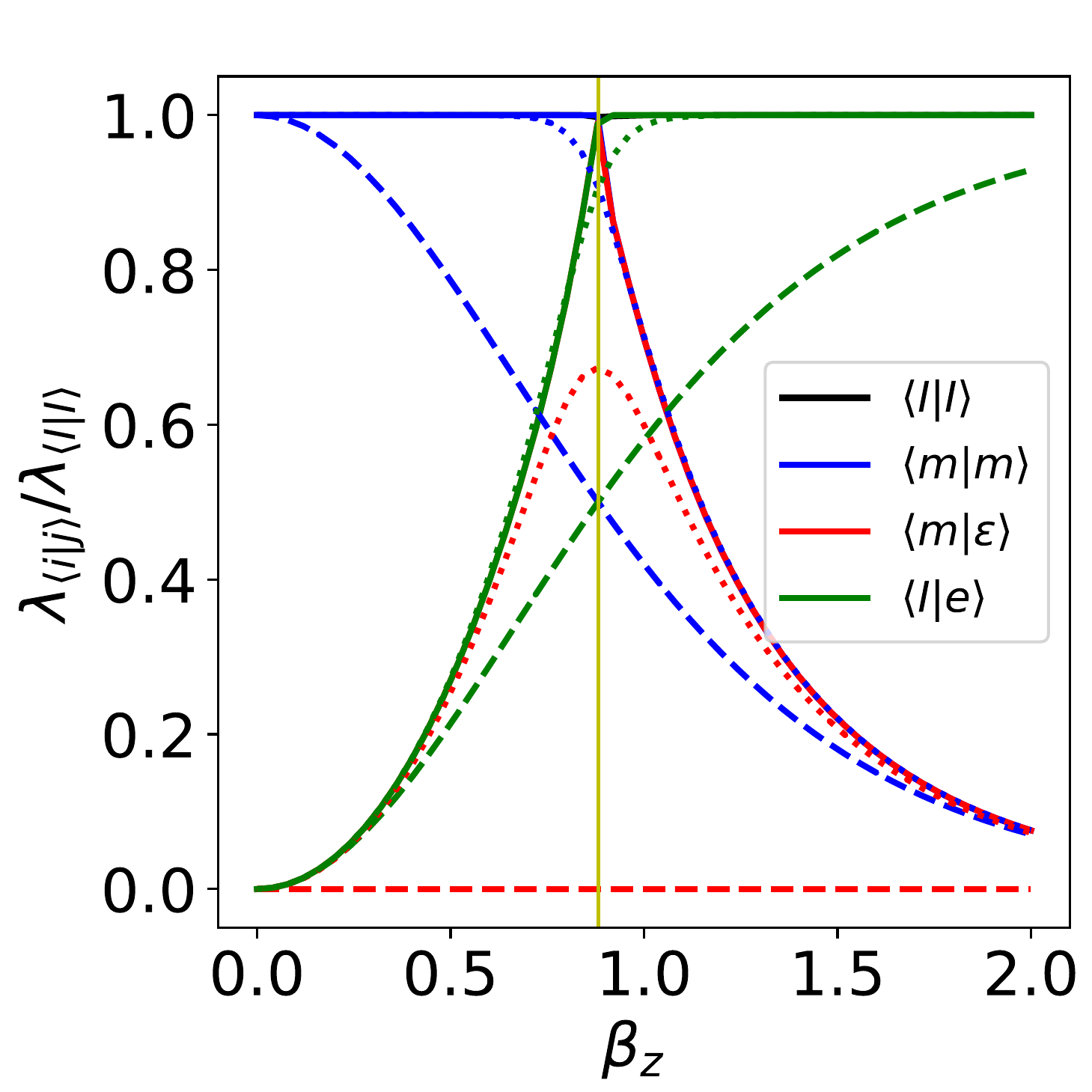}
\caption{Dominant eigenvalues  of the transfer matrices for TC with $L_y = 1$ (dashed lines), $L_y = 8$ (dotted lines), and $L_y = 256$ (solid lines).
} 
\label{fig:TC_data}
\end{figure}

Along the $\beta_x=0$ axis, there exists a  phase transition from  the TO  to the CC phase as shown in Fig.~\ref{fig:TC_data}. 
As we increase $\beta_z$, only the regular and charge difference TMs  [red and blue blocks in Fig.~\ref{fig:double_tensor}(c)] are non-zero, and 
there are only four distinct eigenvalues: $\lambda_{\langle I| I\rangle} = \lambda_{\langle e|e \rangle}$, $\lambda_{\langle m| m\rangle} = \lambda_{\langle \epsilon | \epsilon \rangle}$, $\lambda_{\langle I| e\rangle} =\lambda_{ \langle e| I\rangle}$,  $\lambda_{\langle m| \epsilon\rangle} =\lambda_{ \langle \epsilon| m\rangle}$.
We choose one in each as a representative.
The system exhibits phase transition at $\beta_z = \beta_z^c \approx 0.8814$ [ Fig.~\ref{fig:TC_data}(a)].

We find  $\lambda_{\langle a | b \rangle}$ does not  have significant change for  $L_y > 256$; in the following, we will use $L_y=256$ data to represent the thermodynamic limit.  
For $\beta_z < \beta_z^c$, $\lambda_{\langle m|m\rangle} = \lambda_{\langle I|I\rangle} = 1$ and $\lambda_{\langle I|e\rangle} = \lambda_{\langle m|\epsilon\rangle} <1$. 
 $\lambda_{\langle m|m\rangle} (= \lambda_{\langle \epsilon|\epsilon\rangle})$ and $ \lambda_{\langle I|I\rangle}(=\lambda_{\langle e|e\rangle} )=1$ indicates that $|I\rangle$ and $|m\rangle$ are physically normalizable states, while $\lambda_{\langle I|e\rangle}$ and $ \lambda_{\langle m|\epsilon\rangle}< 1$  suggest that $|I\rangle, |e\rangle, |m\rangle, |\epsilon \rangle$ are four orthogonal states, which is the feature of the four-fold degenerate ground states of the TO phase.
Also, $\lambda_{\langle I|e\rangle} = \lambda_{\langle m|\epsilon\rangle}$ is consistent with Fig.~\ref{fig:double_tensor}(c) that the red blocks and blue blocks should be regarded as the same, respectively.

 For $\beta_z > \beta_z^c$, $\lambda_{\langle I|e\rangle} = \lambda_{\langle I|I\rangle} = 1$ (meaning $|I\rangle = |e\rangle $) and $\lambda_{\langle m|m\rangle} = \lambda_{\langle m|\epsilon\rangle} <1$ (meaning $|m\rangle$ and $|\epsilon\rangle$ are not physically normalizable states). 
 This indicates that the classification in Fig.~\ref{fig:double_tensor}(c) no longer applies; instead, we should identify the blocks in the first (fourth) column as the same.
At $L_y = 1$ and $8$, for $\beta_z < \beta_z^c$, while $\lambda_{\langle I|e\rangle} = \lambda_{\langle m|\epsilon\rangle} $ in the thermodynamic limit, $\lambda_{\langle I|e\rangle} $ is always larger than $\lambda_{\langle m|\epsilon\rangle}$. 
This is in fact due to the $\pi/L$ shift of momentum for $\epsilon$ as mentioned in Sec.~\ref{sec:MES}. 
In particular, {we found that} at  $\beta_z = \beta_z^c$,   $\lambda_{\langle m|m\rangle}$ and $\lambda_{\langle I|e\rangle}$ are always the same regardless of the system size. 
Therefore, we can accurately identify the critical point by using the crossing of $\lambda_{\langle m|m\rangle}$ and $\lambda_{\langle I|e\rangle}$ merely from a single tensor. 
Note that this is only true for $\beta_x = 0$. 
For $\beta_x\ne 0$, if we continuously deform $\beta_z$, the crossing point will shift  as we keep increasing $L_y$. 
This arises from the fact that at in this scenario, not only regular and charge difference blocks but the flux and fermion difference blocks are also non-zero. 
However, even if we fix $\beta_x$ to other values than $0$, after $L > 4$, the crossing point is almost fixed, meaning that we can still identify the critical point using very small system sizes. 

Interestingly, at the critical point $\beta_z = \beta_z^c$, we found that all the leading eigenvalues go to $1$ as the system size increases to $L_y = 256$. 
At first glance, $\lambda_{\langle I|e\rangle} \rightarrow 1$ suggests that the charge anyon is condensed and $\lambda_{\langle m|m\rangle} \rightarrow 1$ implies that the $|m\rangle$ is normalizable. 
However,   $\lambda_{\langle m | \epsilon \rangle} \rightarrow 1$ indicates that the system is gapless, and the above interpretation is ambiguous since the MESs are not well defined in the gapless phase.
Similarly, for the TO to FC transition, only the regular and flux difference  TMs (red and yellow blocks in Fig.~\ref{fig:double_tensor}(b)) are non-zero.
In fact, all the above observations and arguments can be directly adopted to this case once we switch the charge difference to flux difference. 

This idea of using the non-vanishing  TMs to distinguish phases, surprisingly, can also be extended to the Abelian  to non-Abelian TO transition, as we will demonstrate in the next section. 

\section{Kitaev Model on the star lattice}
\label{sec:LG/ SG state }
\begin{figure}[t]
\centering
\includegraphics[width=\linewidth]{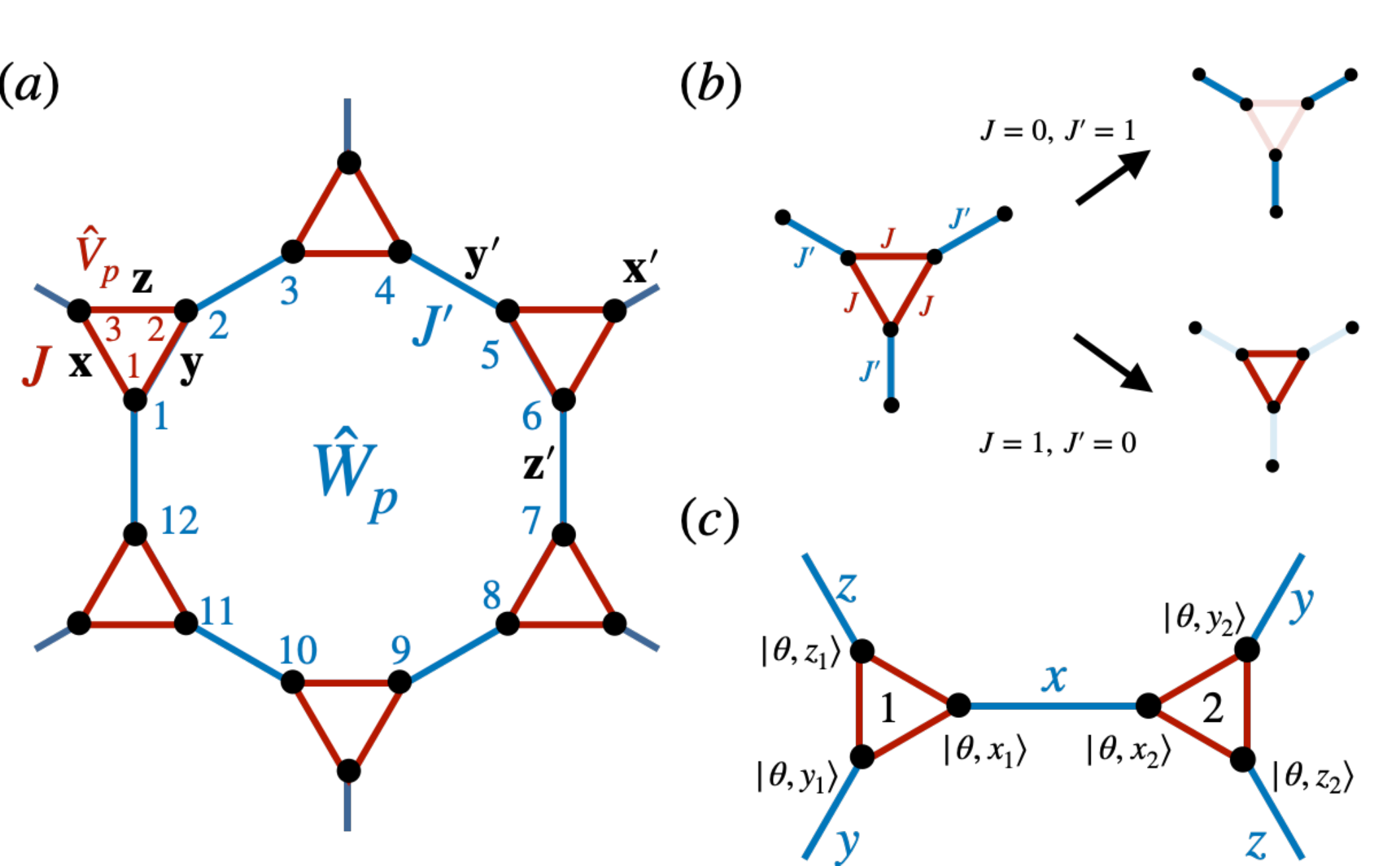}
\caption{ (a) Star lattice with Kitaev-like interactions.  (b) The isolated-dimer and isolated-triangle limits of the Hamiltonian. (c) The initial product state of the LG state.}
\label{fig:star_lattice}
\end{figure}

We extend the ideas developed in the Sec.~\ref{sec:Toric Code} to study the phase transition from an Abelian to a non-Abelian  spin TO.
We will study the quantum phase transition of the  Kitaev model on the star lattice by studying TMs built from  the $\mathbb{Z}_2$-injective  LG and SG states~\cite{spin_one_half,2020-spin-one-kitaev,lee2020anisotropy}.
The Hamiltonian is defined as~\citep{Hong_Ya_2007,KITAEV20062} 

\begin{equation}
H = -J\sum_{{\langle i,j \rangle}_\gamma} S^\gamma_iS^\gamma_j - J'\sum_{\langle ij\rangle \in \gamma'}S^{\gamma'}_i S^{\gamma'}_j
\label{eq:starKitaev}
\end{equation}
where ${\langle i,j \rangle}_\gamma$  and ${\langle i,j \rangle}_\gamma'$  are 
the pairs on  the intratriangle  ($\gamma = x,y,z$) and  the intertriangle  ($\gamma'=x',y',z'$) links connecting site $i $ ans $j$ as shown in Fig.\ref{fig:star_lattice}(a), respectively.  
The Hamiltonian can be block diagonalized by the eigenvalues of two types of flux operators defined on the triangle plaquette $\hat{V}_p = \hat{\sigma}_1^z\hat{\sigma}_2^x\hat{\sigma}^y_3$ and the dodecagon plaquette $\hat{W}_p = \hat{\sigma}_1^x\hat{\sigma}_2^z\hat{\sigma}^y_3...\hat{\sigma}_{12}^y$, where $\hat{\sigma}^i, i = x,y,z$ is the Pauli matrix.
To gain insights into the model, we  consider two extreme limits: the isolated-dimer limit ($J =0,J'=1$) and the isolated-triangle limit ($J =1,J'=0$) [Fig.\ref{fig:star_lattice}(b)].
The perturbative study in Ref.~\cite{KSL_perturbative} shows that in the isolated-dimer limit, while the Hamiltonian does not map exactly onto the standard toric code, the ground state is the same as the toric code on the honeycomb lattice. 
On the other hand, the isolated-triangle limit can be mapped onto the Kitaev honeycomb model at the isotropic point which exhibits a non-Abelian topological order. 
This suggests that there should exist a phase transition between the two phases. 
Exact results shows that the model has two distinct gapped phases: $\mathbb{Z}_2$ topological order when $J'/J >\sqrt{3}$ and non-Abelian CSL with Ising anyon, when   $J'/J <\sqrt{3}$. The latter can be distinguished from the former by its three-fold topological degeneracies which can be labeled using MES basis in Ising anyon: $|I\rangle, |\sigma\rangle, |\epsilon\rangle$.
In both regime, the ground states live in the vortex-free sector $\{W_p = 1, V_p = 1\}$.

\subsection{Loop gas and string gas states}
\label{subsec: LG and SG}
\begin{figure}[tb]
\centering
\includegraphics[width=\linewidth]{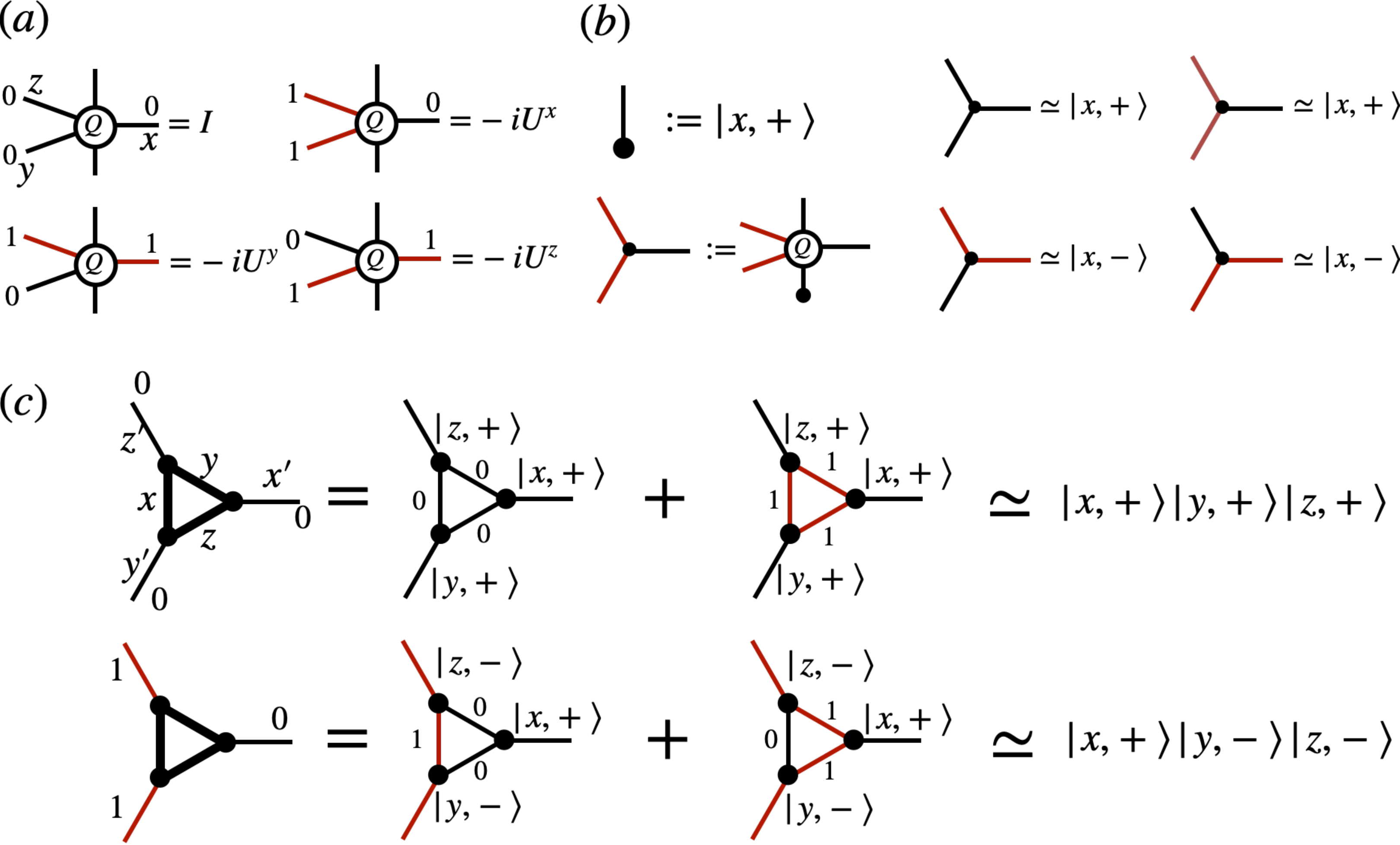}
\caption{(a) Non-zero elements of the LG tensor. Here we denote the the virtual index 0(1) as black(red) leg. (b) The resulting states (up to a phase factor) of LG operators acting on $|x,+\rangle$ for a given set of virtual indices. Similar expression can be derived for the initial states $|y(z), +\rangle$.  (c) The resulting states (up to a phase factor) of LG operators acting on $|\psi\rangle=|x,+\rangle |y,+\rangle |z,+\rangle$ for a given set of virtual indices. Here we use the thick lines to denote that the virtual legs are contracted.} 
\label{fig:LG_star_tnsr}
\end{figure}

Let us first consider an LG operator: $\hat{Q}_{\text{LG}} = \text{tTr}\prod_\alpha {Q}^{ss'}_{i_\alpha j_\alpha k_\alpha}|s\rangle \langle s'| $ with the  non-zero elements of the LG tensor defined as
\begin{equation}
Q_{000} = \mathbb{I}, Q_{011} = -iU^x, Q_{101} = -iU^y, Q_{110} = -iU^z.
\end{equation}
where $U^\gamma = e^{i \pi S^\gamma}, \gamma =x,y,z$ is the $\pi$-rotation operator for a given spin [Fig.~\ref{fig:LG_star_tnsr}(a)] ~\cite{lee2020anisotropy}. 
 By construction, the LG tensor is invariant under the global $\mathbb{Z}_2$ symmetry on the virtual indices: $Q(u_g \otimes u_g \otimes u_g) = Q $ with $u_g = \mathbb{I}, {\sigma}^z$.
Therefore, applying $\hat{Q}_{\text{LG}}$ on any injective PEPS yields a $\mathbb{Z}_2$-injective PEPS. 
 $\hat{Q}_{\text{LG}}$ is a projector to the vortex-free space such that $\hat{W}_p\hat{Q}_{\text{LG}} = \hat{Q}_{\text{LG}}\hat{W}_p = \hat{Q}_{\text{LG}}$, $\hat{V}_p\hat{Q}_{\text{LG}} = \hat{Q}_{\text{LG}}\hat{V}_p = \hat{Q}_{\text{LG}}$.
Another interesting property of $\hat{Q}_{\text{LG}}$ is that the creation of flux anyon pair discussed in Sec.\ref{sec:Z2-injective} now corresponds to two vortices $W_p = -1$ at the endpoint of the string $u_g^{\otimes L}$ \citep{spin_one_half}.

The LG state can then be obtained by applying $\hat{Q}_{\text{LG}}$ on an initial product state $|\Psi(\theta) \rangle = \otimes_\alpha |\psi_\alpha (\theta) \rangle$ where $\alpha$ is the sites index for a given triangular plaquette and $|\psi_\alpha (\theta)\rangle =  |\theta, x_\alpha \rangle  |\theta, y_\alpha \rangle  |\theta, z_\alpha \rangle$ [Fig.~\ref{fig:star_lattice}(c)]. 
 The magnetic state $ |\theta, \gamma_\alpha \rangle$ satisfies
\begin{equation}
\langle \theta, \gamma| \sigma^{\gamma'}|\theta , \gamma \rangle = \delta_{\gamma' \gamma} \cos \theta + (1-\delta_{\gamma' \gamma})\frac{\sin \theta}{\sqrt{2}}, 
\end{equation}
where $\theta$ is a tunable parameter and $\gamma, \gamma'=x,y,z$.
To simplify the notation, we follow the convention in Ref.~\cite{non-AbelianTO_2020} to parametrize the Hamiltonian $H = H(\phi)$ with $J' = \sin(\phi), \ J = \cos(\phi)$. 
The ground state for a given Hamiltonian $H(\phi)$ can  then be obtained by variationally optimizing the parameter $\theta$ to find the lowest energy.

To gain more insights, we first consider two limits where the LG state is the exact ground state.  
In the isolated-dimer limit,  $H(\phi = \pi/2)$, the  ground state degeneracy is exponentially large with the system size, and one of the ground state basis state is a product state $|\Psi \rangle = \otimes_\alpha |\psi_\alpha \rangle$ with $|\psi\rangle = \big( |x,+\rangle | y,+\rangle |z, + \rangle \big)$, where $|\gamma,\pm\rangle$ is the eigenvector of $\sigma^\gamma$ with $\pm 1$ eigenvalues for $\gamma = x,y,z$. 
This basis is the initial product state for the LG states with $\theta = 0$ since  $\langle \gamma| \sigma^{\gamma'}| \gamma \rangle = \delta_{\gamma' \gamma}$, and thus one can identify $|\gamma,+\rangle = |\theta = 0, \gamma \rangle$.
If we slightly deviate from  $\phi=\pi/2 $, the state is no longer the ground state. 
However, we expect the ground state of the model to live in the vortex-free sector, and  
we can apply $\hat{Q}_{\text{LG}}$ to project it back to the vortex-free space,  again giving the LG state at $\theta = 0$.

Using the fact that $Q_{011}, Q_{101}, Q_{110}$ are  the $\pi$-rotation operator (up to a phase factor) around the $x, y, z$-axes, one can derive the resulting state of the LG operator on $|\theta = 0, \gamma \rangle$ (i.e., $|\gamma,+\rangle$) as in  Fig.~\ref{fig:LG_star_tnsr}(b).
Now we can combine three different initial states together to form a triangular product state: $|x,+\rangle |y,+ \rangle |z,+ \rangle $ [Fig. ~\ref{fig:LG_star_tnsr} (c)].
For a given set of virtual indices,   the triangular LG state can be written exactly as the sum of two terms due to the $\mathbb{Z}_2$-invariance. 
Using the relation in Fig.~\ref{fig:LG_star_tnsr}(b), one can find that the two terms are exactly the same, as shown in Fig.~\ref{fig:LG_star_tnsr}(c). 
Furthermore, the physical state with different virtual indices, e.g., $|y,+\rangle,\ |y,- \rangle$ and $|z,+\rangle,\ |z,- \rangle$ in Fig.~\ref{fig:LG_star_tnsr}(c), are orthogonal.
This property, combining with the fact the the tensor is $\mathbb{ Z}_2$-invariant, guarantees that the LG state with $\theta = 0$ is $\mathbb{ Z}_2$-isometric~\citep{2011_Norbert_Ginjective}. 
Interestingly, the $\mathbb{ Z}_2$-isometry 
{shows}
 that LG state at $\theta = 0$ is the RG fixed point of $\mathbb{Z}_2$ topological order. 
This is consistent with the fact that in the  isolated-dimer limit, the ground state is the same as the that for the TC on the honeycomb lattice ~\citep{KSL_perturbative}. 
Note that since SG states is an extension for LG states, it is also $\mathbb{ Z}_2$-isometric at $\phi = \pi/2$. 
The property of $\mathbb{ Z}_2 $-isometry allows us to view the toric code with string tension (discussed in Sec.~\ref{sec:Toric Code}) and LG, SG with different parameters on the same footing.

On the other hand, in the isolated-triangle limit, $H(\phi=0)$, the LG state is the exact ground state~\citep{non-AbelianTO_2020,KSL_perturbative}.
However, for $0<\phi<\pi$, the energy of the optimized  LG state is higher than the exact value~\cite{non-AbelianTO_2020,Hong_Ya_2007}. 
Therefore, instead of using  the optimized LG state for  a specific Hamiltonian $H(\phi)$, in the following we tune the parameter $\theta$ in the  LG state to study its property. 

Similarly, for the SG state, we  introduce the dimer gas (DG) operator $\hat{R}_{\text{DG}}(\alpha, \beta) = \text{tTr}\prod_\gamma {R}^{ss'}_{i_\gamma j_\gamma k_\gamma}(\alpha, \beta)|s\rangle \langle s'| $ with a DG tensor
\begin{equation}
R^{ss'}_{ijk}(\alpha, \beta) = \zeta_{ijk}(\alpha, \beta)[(\sigma^x)^i(\sigma^y)^j(\sigma^z)^k]_{ss'}.
\label{eq:SG}
\end{equation}
where
\begin{equation}
  \zeta_{ijk}(\alpha, \beta)=
    \begin{cases}
      \cos{\beta} & \text{if    } i+j+k = 0 \ \text{(mod 2)}\\
      \sin{\alpha} & \text{if    } i+j+k = 1 \ \text{(mod 2)}
    \end{cases}.
\end{equation}
The SG state can be constructed as $|\psi_\text{SG} (\alpha, \beta)  \rangle = \hat{Q}_\text{LG}\hat{R}_\text{DG}(\alpha, \beta) |\psi(\theta = \tan^{-1}\sqrt2)\rangle$~\cite{non-AbelianTO_2020}.
Since the SG state yields quite accurate ground state energy for  the star-lattice Kitaev model, instead of considering  the SG state as a two-parameter family of $\mathbb{ Z}_2 $-injective tensor, we label them using the Hamiltonian parameter $\phi$ instead.

\subsection{Overlap of minimally entangled states}
\label{subsec:overlap of mes}
In the following, we  study the topological properties of the LG and SG states by computing the overlap of MESs, which corresponds to  the dominant eigenvalues of the TM blocks, $\lambda_{\langle a | b \rangle}$. 

As we continuously change the parameter for both the LG and SG states, we find that only the regular and the  TMs measuring the fermion difference (red and green blocks in Fig.~\ref{fig:double_tensor}(c)) are non-zero.
This means that the charge and flux anyons are static while the fermion is dispersive, consistent with the exact results that only the Majorana fermion has dynamics \citep{KITAEV20062,Hong_Ya_2007}.
In fact, we find that any arbitrary PEPS state applied by the spin-1/2 LG projector always possess static charge and flux, suggesting that the LG projector is indeed a suitable operator to approximate the spin-1/2 Kitaev spin liquids.
On the other hand, the very recent work showed that any arbitrary PEPS state applied by the spin-1 LG projector harbors dispersive charge excitations, yet the flux and fermions are static~\citep{spin1}. 
Theoretical studies about the distinct excitation property of the integer and half-integer Kitaev spin liquids arising from the different sign structures of the LG projectors are worth investigating in the future.
Also, we find that there are only four distinct eigenvalues, $\lambda_{\langle I | I\rangle }= \lambda_{\langle \epsilon| \epsilon\rangle} $, $\lambda_{\langle m| m\rangle} = \lambda_{\langle e| e\rangle} $, $ \lambda_{\langle m| e\rangle} = \lambda_{\langle e| m\rangle} $, and $\lambda_{\langle I| \epsilon\rangle} = \lambda_{\langle \epsilon| I\rangle} $, while others are zero. 
Therefore, in each category we choose one as the representative.  
Note that since  $\lambda_{\langle I | I \rangle} $ is always the largest, we normalize it to 1. 

\begin{figure}[t]
\centering
\includegraphics[width=\linewidth]{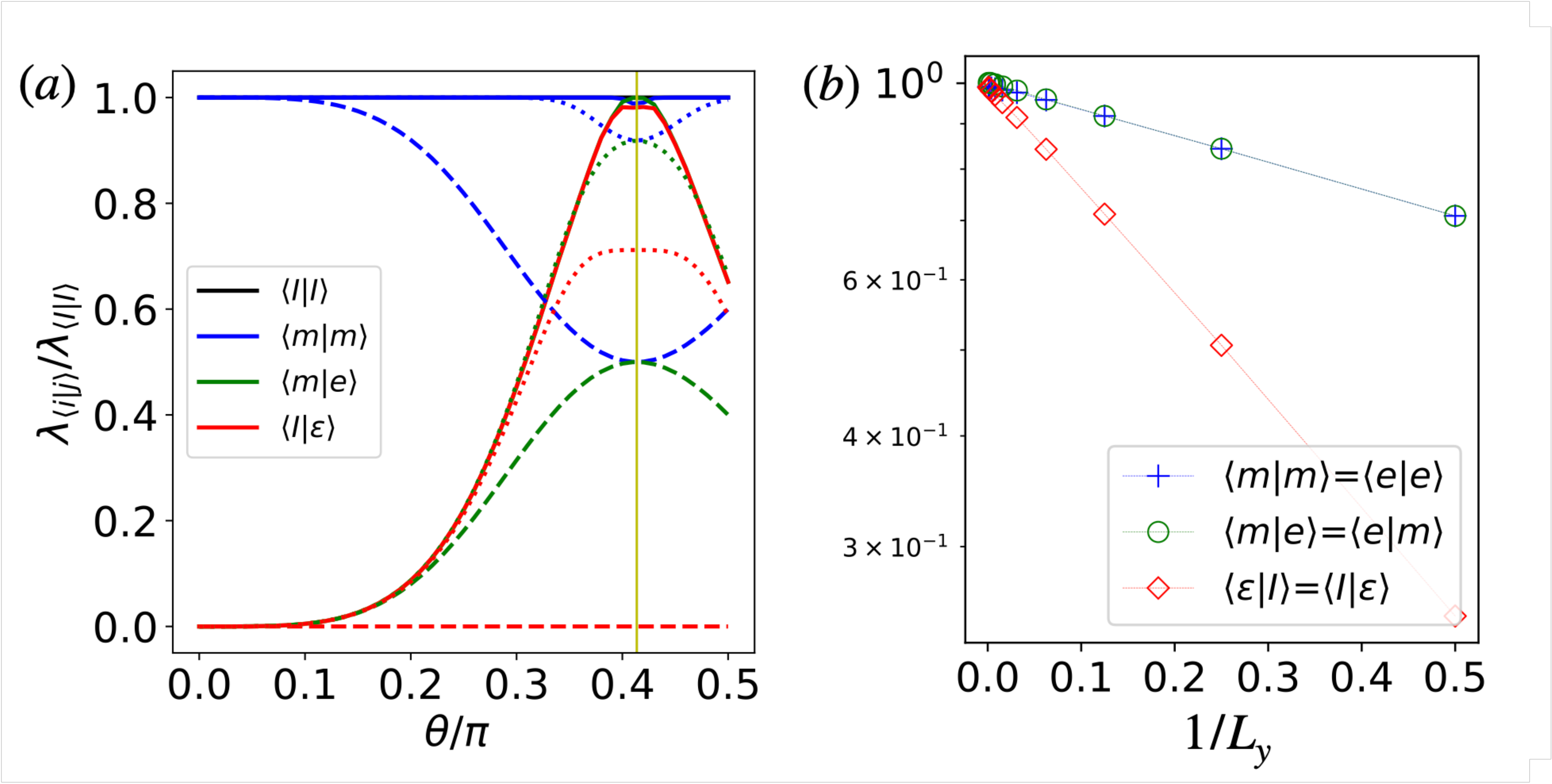}
\caption{  (a)  Dominant eigenvalues of the transfer matrix for LG states with $L_y$ = 1 (dashed lines), $L_y$ = 8 (dotted lines), and $L_y$ = 256 (solid lines).  (b)  Dominant eigenvalues of the transfer matrix  for $\theta=\theta_c$ with $D_\text{cut} = 80$. }
\label{fig:LG_data}
\end{figure}

Figure~\ref{fig:LG_data} shows the overlaps of the LG states for $L_y = 1, 8$, and $256$. 
For $L_y = 1, 8$, we can see that as $\theta$ increases from $0$ to $\theta_c = \cos^{-1}{(2-\sqrt{3})}$, $\lambda_{\langle m| m\rangle}$ gradually decreases and $\lambda_{\langle m| e\rangle}$ gradually increases. 
At $\theta = \theta_c$, these two eigenvalues are exactly the same, meaning that we have reached the transition point. 

However, if we keep increasing $\theta$, the $\lambda_{\langle m| m\rangle}$ and $\lambda_{\langle m| e\rangle}$ do not cross. 
Unlike the level crossing for the charge or flux condensation transition in the toric code with string tension (see Sec.~\ref{sec:Toric Code}), here the dominant eigenvalues for the TM blocks only touch, indicating that $\theta < \theta_c$ and $\theta > \theta_c$ are in the same phase.
At  system size $L_y = 256$, we find that $\lambda_{\langle m| e\rangle} = \lambda_{\langle I| \epsilon\rangle} < 1$ for $\theta \neq \theta_c$. 
This means that $\mathbb{T}_{\langle m| e\rangle} $ and $\mathbb{T}_{\langle I| \epsilon\rangle}$ are the same, and can be grouped with the same (green) color in Fig.~\ref{fig:double_tensor}(c). 
Therefore, all the LG states live in the topologically ordered phase, except at $\theta = \theta_c$.

At the transition point $\theta = \theta_c$, all $\lambda$'s approach 1 as the size increases. 
Similar to the transition points studied in Sec.~\ref{sec:Toric Code}, $\lambda_{\langle I| \epsilon\rangle} \to 1$ indicates a gapless excitation.
On the other hand, we find that $\lambda_{\langle m| e\rangle}$ is always equal to $\lambda_{\langle m| m\rangle}$ and $ \lambda_{\langle e| e\rangle}$ regardless of the system size (see Fig.~\ref{fig:LG_data}(b)). 
It suggests that we can  identify $|e\rangle$ and $|m\rangle$ even though the system is gapless.
This argument is supported by the calculation of topological entropy $\gamma$. 
We find that both $|e\rangle$ and $|m\rangle$ can yield exactly the same $\gamma = 1/2 \times \ln(2)$ at $\theta = \theta_c$, while $|I\rangle$ and $|\epsilon\rangle$ give $\gamma = \ln 2$. 
Also, by identifying  $|e\rangle$ with $|m\rangle$, we can explain the three-fold degeneracy of the ground state with the correct total quantum dimension $D = \sqrt{1^2+1^2+(\sqrt{2})^2} = 2$.

\begin{figure}[t]
 \centering
\includegraphics[width=\linewidth]{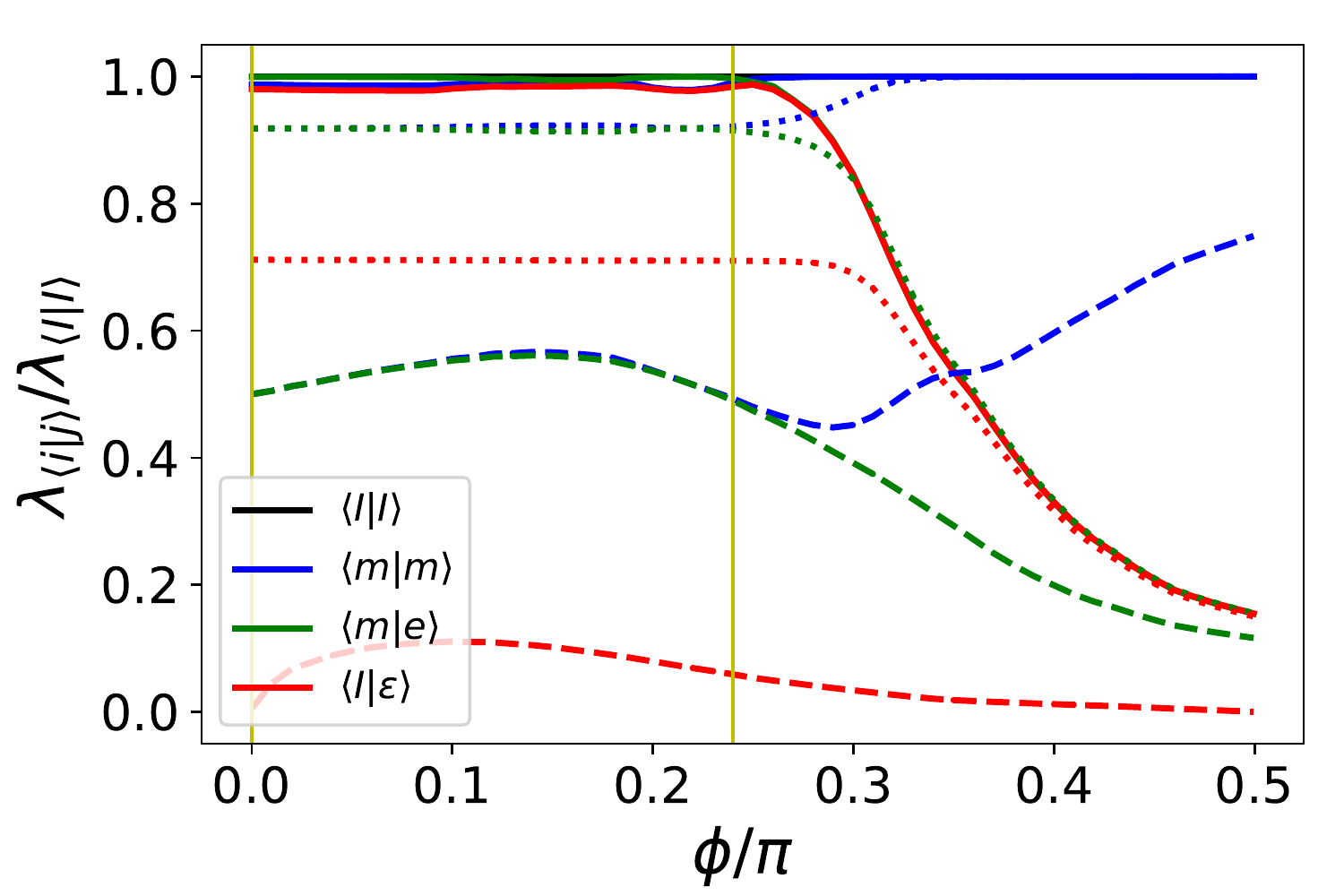}
\caption{The dominant eigenvalues of the transfer matrices for the SG states with $L_y = 1$ (dashed lines), $L_y = 8$ (dotted lines), and $L_y = 256$ (solid lines).} 
\label{HOTRG_SG_DC64}
\end{figure}

To obtain the SG state for each $\phi$, we optimize the free parameters $\alpha$ and $\beta$ in Eq.~\eqref{eq:SG} to obtain the variational ground state of Eq.~\eqref{eq:starKitaev}.
Figure~\ref{HOTRG_SG_DC64} shows the dominant eigenvalues of the transfer matrices constructed from the SG state. 
Recall that the ground state of the star lattice  Kitaev is an Abelian spin liquid at $\pi/3<\phi < \pi$, and non-Abelian at $0<\phi<\pi/3$.
For $\phi = \pi$, the SG state is $\mathbb{ Z}_2 $-isometric just like the LG state at $\theta = 0$. 
For $L_y = 1$, as $\phi$ decreases from $\phi = \pi$ to $\phi \approx 0.24\pi$, $\lambda_{\langle m| m\rangle}$ gradually decreases and $\lambda_{\langle m| e\rangle}$ gradually increases. 
At $\phi \approx 0.24\pi$, these two eigenvalues become identical. 
Different from the transition in Sec.~\ref{sec:Toric Code}, if we keep decreasing $\phi$, both $\lambda_{\langle m| m\rangle}$ and $\lambda_{\langle m| e\rangle}$ increase together.
In fact, $\lambda_{\langle m| e\rangle}$ should never become lager than $\lambda_{\langle m| m\rangle}$ since the dominant eigenvalues of the regular transfer matrix will always be larger than other blocks for the norm of the states to stay positive. 
However, the trend for $\phi < 0.24\pi$ is also different from the LG case. 
The increase of both $\lambda_{\langle m| m\rangle}$ and $\lambda_{\langle m| e\rangle}$ strongly suggest that $|e\rangle$ becomes $|m\rangle$ in that regime. 
This is also consistent with the topological entanglement  entropy of $|m\rangle$ becomes $1/2\log(2)$ at $\phi < 0.24\pi$~\cite{non-AbelianTO_2020}. 
As $\phi \rightarrow 0$, both $\lambda_{\langle m| m\rangle}$ and $\lambda_{\langle m| e\rangle}$ begin to decrease to the same point as $\phi \approx 0.24 \pi$, suggesting that we have two transition points at $\phi = 0$ and $\phi = 0.24 \pi$. 

As we further increase the circumference to $L_y = 256$, all $\lambda s \rightarrow 1$ for $\phi < 0.24\pi$. 
In particular, $\lambda_{\langle I| \epsilon \rangle} \rightarrow 1$  suggests that the parent Hamiltonian of the SG states is gapless in this regime.
This result is compatible with  the no-go theorem~\cite{2015_no_go_theorem} that the parent Hamiltonian of  a chiral PEPS is gapless, contrary to the claim in Ref.~\cite{non-AbelianTO_2020}.
However, as shown in Ref.~\cite{2013_PEPS_chiral_TO}, there might exist other non-frustration-free gapped Hamiltonian, in our case the Kitaev star lattice Hamiltonian, whose ground state can  be well approximated by the SG states. 

\subsection{Transfer Matrix Spectrum}
\label{subsec:tm_spectrum}

\begin{figure}[tb]
\centering
\includegraphics[width=\linewidth]{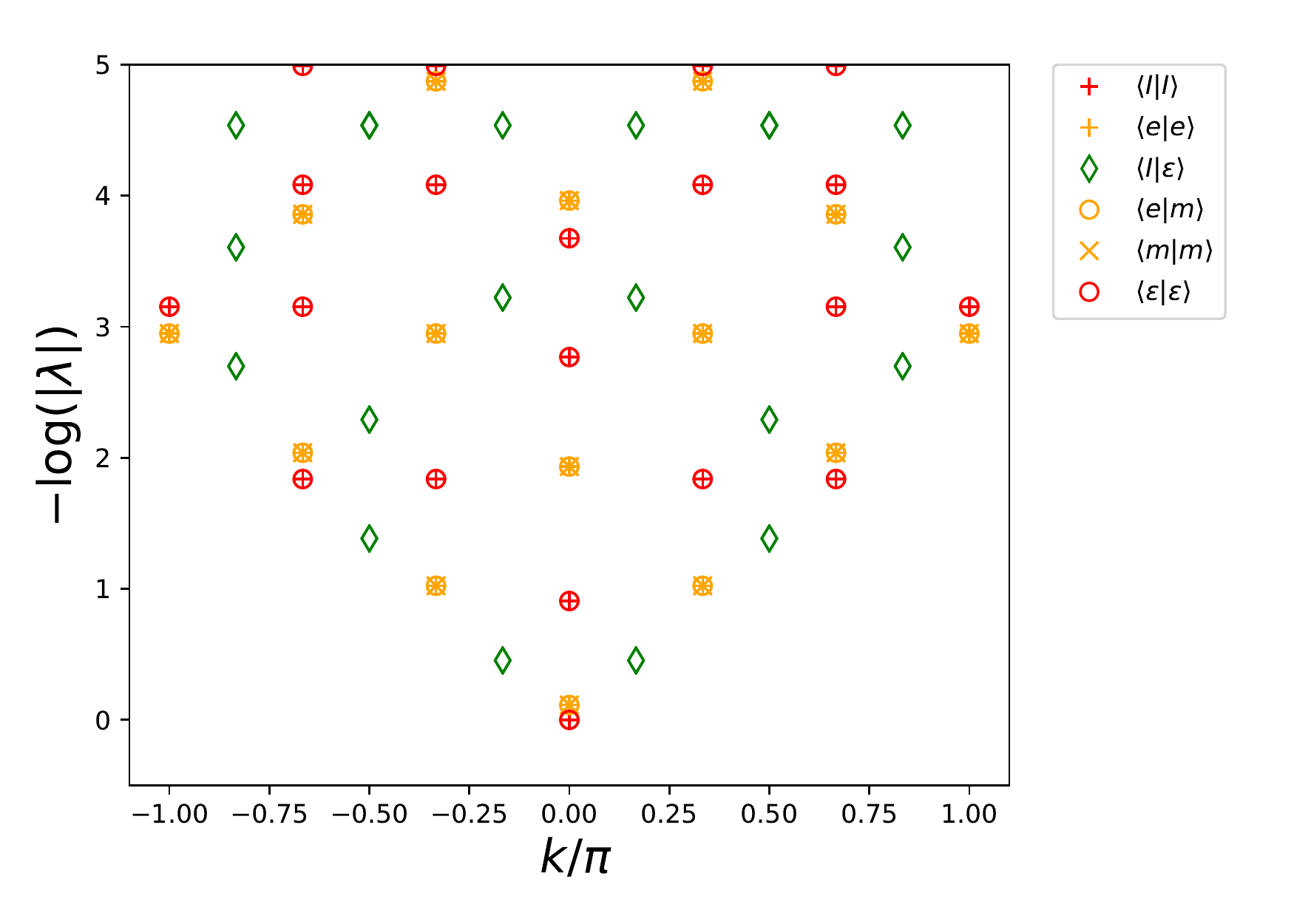}
\caption{ Transfer matrix spectrum of LG at $\theta = \theta_c$ with $L_y = 6$.} 
\label{LG_thetaC_L6_NA}
\end{figure}

To further support our claim that the $|m\rangle$ and $|e\rangle$ are identical, we present
the full spectrum of the transfer matrices labeled by the momentum quantum numbers. 
Figure~\ref{LG_thetaC_L6_NA} shows the TM spectrum (minus logarithms of the eigenvalues) for LG state at $\theta = \theta_c$. 
One can observe that not only the dominant eigenvalues match $\lambda_{\langle e|e\rangle} = \lambda_{\langle m|m\rangle} = \lambda_{\langle m|e\rangle}$, but their full spectra also match. 
This means that the two MES $|m\rangle$ and $|e\rangle$ are exactly the same state. 
In contrast, while in the thermodynamic limit $\lambda_{\langle I|I\rangle}  = \lambda_{\langle I|\epsilon\rangle} $, their spectra are always different. 
This strongly suggests that $\lambda_{\langle e|e\rangle} = \lambda_{\langle m|m\rangle} = \lambda_{\langle m|e\rangle}$ is due to the degeneracy of the state while  $\lambda_{\langle I|I\rangle}  = \lambda_{\langle I|\epsilon\rangle} $ is due to the mode softening.

Recall the discussion in Sec.~\ref{sec:Toric Code}, once we drive the $\mathbb{Z}_2$-injective wave functions from the TO phase to the CC  phase, the original MES basis is no longer the appropriate basis. 
The $|e\rangle$ becomes  exactly the same as $|I\rangle$, and the $|m\rangle$ is not  a physical normalizable state. 
Similarly, for the LG and SG states in the non-Abelian regime, there exist no charge and flux anyons anymore. 
Combining with the calculation of entanglement entropy, we can  regard the charge and flux transmute into $\sigma$ anyon for $\phi < 0.24\pi$.
 For the honeycomb Kitaev model, the ground state is Abelian when $|J_x| \geq |J_y|+|J_z|$, and the extreme limit can be directly mapped to the TC~\cite{Kitaev2006}. 
There, the charge and flux lives in alternating rows of plaquettes. 
 On the other hand, in the non-Abelian phase, all the plaquettes should be regarded equal and the vortex excitation is the  $\sigma-$anyon.
Since the Abelian and non-Abelian limit can be respectively mapped to the TC and the isotropic  honeycomb  Kitaev model, there exists a critical point where $|e\rangle$ and $|m\rangle$ transmute into $|\sigma\rangle$.
However, the string of $u_g$ and the charge operator $R_\alpha$ encode the fusion and braiding rules for the $\mathbb{Z}_2$ topological order, which can not describe the  non-Abelian case. 
This is the limitation of the $\mathbb{Z}_2$ classification built in the $\mathbb{Z}_2$-injective PEPS.
As we  show in Sec.~\ref{subsec:overlap of mes}, the parent Hamiltonian is  gapless for $\phi < 0.24\pi$, which does not support gapped $\mathbb{Z}_2$ anyons such as $|e\rangle$ and $|m\rangle$.
Within the constraint of the  $\mathbb{Z}_2$-injective PEPS, the best approximate wave function for a non-Abelian CSL is to make $|e\rangle$ and $|m\rangle$ identical.
On the other hand, the   star lattice  Kitaev model is \textit{not} the parent Hamiltonian of the SG states  and  excitations can be gapped for $\phi < 0.24\pi$. 
This means that the anyonic excitation  described by a $\mathbb{Z}_2$-injective PEPS may not be the true excitation of the model. 
Nevertheless, one can create an excitation by the string action $u_g^{\otimes L} = (\hat{\sigma}^z)^{\otimes L}$, which will create a vortex pair with $W_p = -1$ at the endpoints of the string. 

\section{Discussion and outlook}
\label{sec:Conclusion}
Now we have  a unified picture to describe the transitions from TO to CC, FC, and non-Abelian phases in terms of the 16 blocks of the TM. 
The TO to CC transition can be detected when  the blue blocks in Fig.~\ref{fig:double_tensor}(c)  become distinct. 
To be more specific, as $|I\rangle$ and $|e\rangle$ become the same state, $|m\rangle$ and $|\epsilon\rangle$ are confined, and thus $\langle I|e\rangle$ and $\langle m| \epsilon \rangle$ are different. 
Similarly, the emergence of the FC(non-Abelian) phases can be observed as the yellow(green) blocks become distinct.
Different from the CC and FC case, the non-Abelian case is not accompanied with the confinement of other particles. 
Since the parity even sector of a $\mathbb{ Z}_2$-invariant tensor is always non-zero, the vacuum state $|I\rangle$ is always normalizable. 
In addition, due to the fact that no other MES can become $|\epsilon \rangle$, we conclude that $\mathbb{ Z}_2$-injective tensors can only detect three types  of anyon transitions from identifying the MESs:   $|e\rangle = |I\rangle$, $|m\rangle = |I\rangle$, or  $|e\rangle = |m\rangle$. 
However, there exist other types of  topological phase transitions beyond this scheme. 
For instance, in Ref.~\cite{Zhang_2019}, it is shown that the self-dual phase transition point of the TC wave function corresponds to  the Kramers-Wannier duality of the Ashkin-Teller model, where none of the MESs become identical.

In the current work, we use the $\mathbb{Z}_2$-injective PEPS as an example to identify and classify topological phase transitions out the the $\mathbb{Z}_2$ TO.
However, this scheme can be easily generalized to $G$-injective and MPO-injective PEPS~\cite{BULTINCK2017183}.
The method developed here for computing dominant eigenvalues of TMs is a powerful tool to determine whether the system undergoes a phase transition when the PEPS tensor acquires virtual symmetry.
The low computation cost of this HOTRG-inspired method makes it suitable to perform finite-size scaling analysis, which can be used to extract  scaling dimensions at the critical point. 
Further studies along these directions are worth pursuing.

\acknowledgements

This work is partially supported by the Ministry of Science and Technology (MOST) of Taiwan under grants 
No.~108-2112-M-002-020-MY3, 
No.~107-2112-M-002-016-MY3, and 
No.~108-2112-M-029-006-MY3.
We thank J. Genzor for collaboration on related work.

\appendix

\section{Computing the Dominant Eigenvalues of Transfer Matrices on the Long Cylinder}
\label{sec: tnTM}

As described in the main text, the dominant eigenvalues of 16 blocks of TM are essential to distinguish different phases. To compute the TMs on the long cylinder, we merge the tensors along $y$-direction using a HOTRG-inspired method and preserve the gauge symmetry which extends the idea in Ref.~\cite{GSPRG_2014}.  Our approach can be described in the following steps.

(1) {\it Creating the double tensor\/}.
Starting from a local tensor $A$,  we contract physical indices of   $A$ and its adjoint $A^*$  to form the double tensor  $\mathbb{ E} $,
 $\mathbb{ E}  \equiv \sum_s ( A^s_{i,j,k,l} ) \times ( A^s_{i',j',k',l'} )^* $.

\begin{figure}[bp]
 \centering
\includegraphics[width=\linewidth]{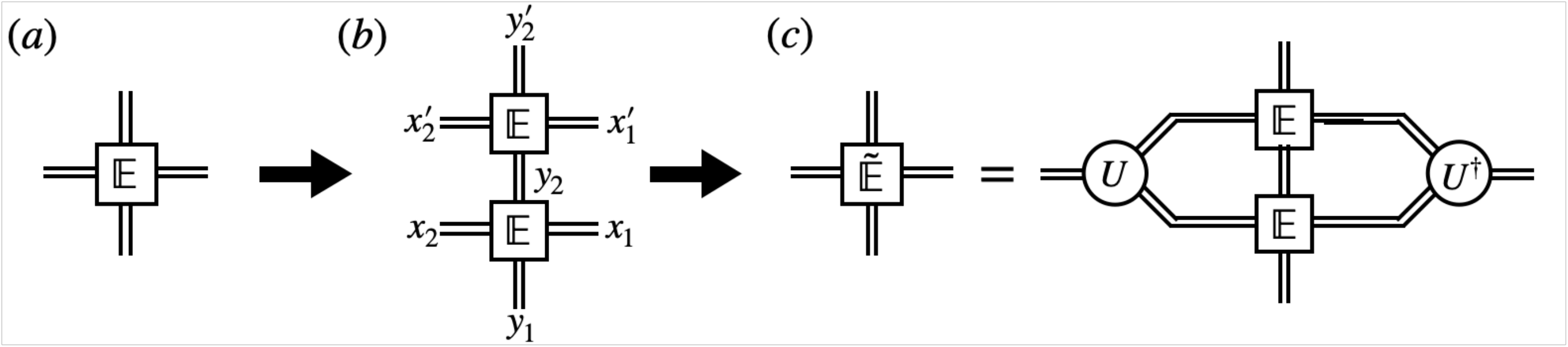}
\caption{(a) Start from a one-site double tensor. (b) Merge two double tensors to form a new rank-6 tensor. (c) Apply appropriate isometry $U$ which truncates the bond dimension}  
\label{fig:coarse_grain}
\end{figure}

\smallskip (2) {\it Coarse graining tensors with preserved gauge symmetry\/}.
Two sites are merged into a single site along $y$-direction, generating a rank-6 tensor  $\mathbb{ E'} =  \sum_{y_2} \mathbb{ E}_{x_1,y_1,x_2,y_2}  \mathbb{ E}_{x'_1,y_2,x'_2,y'_2} $, 
where the indices of $\mathbb{ E} $ start on the right and go around clockwise to the top. 
This can be regarded as a rank-4 tensor by formally grouping the two indices ($x_1, x'_1$) on the right to one index, and similarly the two on the left to another. 
The bond dimension of tensor  $\mathbb{ E'}$  along the cylinder direction is the square of the original bond dimension of tensor $\mathbb{ E}$. 
Applying an appropriate isometry $U$ truncates the size of these squared bond dimensions to a fixed number, say, $D_\text{cut}$, and a truncated tensor $ \tilde { \mathbb{ E}}$ can be obtained [Fig.~\ref{fig:coarse_grain}].

To determine the isometry $U$, we note that the double tensor $\mathbb{E}$ enjoys the $\mathbb{Z}_2 \times \mathbb{Z}_2$ symmetry inheriting from the $\mathbb{Z}_2$-invariant tensor, which can be written in the block-diagonal form: $\mathbb{E} = \oplus_{\alpha, \alpha'} \mathbb{E}^{\alpha'}_\alpha$ where $\alpha, \alpha' = \pm$.
When two sites are merged into a single site, we identify  $(\alpha_1,\alpha_2) = (+,+), (-,-)$ as $\alpha = +$ and $(\alpha_1,\alpha_2) = (+,-),(-,+)$ as $\alpha = -$ on both the bra and ket layers. 
The isometry $U^{\alpha'}_\alpha $ for each block can be obtained by performing eigenvalue decomposition to the following tensor:

\begin{multline}
(M^{\alpha'}_\alpha)_{x_1  x'_3,  x_2  x'_3} 
=  \sum_{x_2,x_2',y_1,y_2} \\
(\mathbb{E'}^{\alpha'}_\alpha)_{x_1,x'_1,y_1,x_2,x'_2,y_1} 
(\mathbb{ E'}^{\alpha'}_\alpha)^*_{x_2,x'_2,y_2,x_3,x'_3,y_2}.   
\end{multline}

One can then apply the isometry $U =  \oplus_{\alpha, \alpha'} U^{\alpha'}_\alpha $ onto $\mathbb{E'}$ to generate the truncated tensor $\tilde{\mathbb{E}}$ preserving $\mathbb{Z}_2 \times \mathbb{Z}_2$ symmetry.

After $p$ steps of this iteration, the final tensor represents a chain of  $2^p$ tensors that preserves the gauge symmetry.

\smallskip (3) {\it Inserting the string operator and taking the trace\/}.
Finally, we could insert the string operator $S^{g'}_g = u_{g'} \otimes u_g$ and take the tensorial trace along the $y$-direction for each block $\mathbb{E^{\alpha'}_\alpha}$. The combination of $(g,g') = I,Z$ and $(\alpha, \alpha') = +,-$ then gives 16 blocks of TM, where their dominant eigenvalues can be computed using the standard Krylov method.

\bibliography{bibs}

\end{document}